\documentclass[]{article}
\usepackage{amssymb,latexsym,amsmath}     
\usepackage{authblk} 
\usepackage{enumerate}
\usepackage{mathtools}
\usepackage[title]{appendix}
\usepackage[utf8]{inputenc}
\usepackage{euscript}
\usepackage{dsfont}
\usepackage{cool}
\usepackage{url}
\usepackage{hyperref}
\usepackage{xcolor}
\DeclareFontFamily{OT1}{pzc}{}
\DeclareFontShape{OT1}{pzc}{m}{it}{<-> s * [0.900] pzcmi7t}{}
\DeclareMathAlphabet{\mathpzc}{OT1}{pzc}{m}{it}

\begin{document}
\title{Analogue gravity and its scientific confirmatory role}

%\author[1]{ Saeed Masoumi \thanks{s\_masoumi@sbu.ac.ir} }
\author{ Mojtaba Shahbazi \thanks{mojtaba.shahbazi@modares.ac.ir} }

%\affil[1]{Institute for Science and Technology Studies, Shahid Beheshti University, Tehran,
%	Iran}
\affil{Department of Physics, Faculty of Science, Tarbiat Modares University, Tehran, Iran}

\maketitle

%----------------------------------------------------------------
\begin{abstract}
Empirical confirmation in some areas of physics is obscure; for example in Hawking radiation. However, the analogue gravity can simulate these phenomena in condensed matter systems. That is an important question whether the observation of these phenomena in the condensed matter systems can be confirmatory of the original phenomenon or not. In this work we answer affirmatively to this question via structuralism.
\end{abstract}

%--------------------------------------------------------------------
\section{Introduction}
Analogue gravity (AG) teaches us how a field theory in a curved background could be simulated by a sound wave in a fluid, generally a perturbation in a condensed matter system. Especially, this method provides a practical way to simulate a physical phenomenon which is not accessible, say, a black hole. One of the great achievements of the analogue gravity is detection of analogue of the Hawking radiation in a water tank [Weinfurtner et.al, 2011] and Bose-Einstein condensation [Steinhauer, 2016]. The detection of the original Hawking radiation requires two things: one, an evaporating black hole which the nearest one is very far away from Earth and the second, the technology of detectors because the Hawking radiation is too weak particularly when it is compared with the cosmic microwave background, then the required detectors are beyond our cutting-edge technology. In this vein, the Hawking radiation is already a far reached physical phenomenon. The other inaccessible example  is string theory. The length of fundamental string is of order $10^{-33}cm$, then to detect a fundamental string requires an accelerator of energy order $10^{16}TeV$ , if compared to LHC which is of order $7 TeV$, it seems that the detection of a fundamental string is due to too-distant future. Dealing with these theories and phenomena prompts speculation that if these theories are experimentally confirmed or not? How does one experimentally confirm them? What are the confirmatory criteria in these cases?

As a first instinct, if one can simulate physical phenomena, they can take the advantage of them, it means that if a black hole acts as a natural quantum computer, then an analogue system which simulates a black hole can act as an analogue of a quantum computer, it means the dynamics of the two are the same. Nonetheless, it is believed that the analogue gravity takes the role of confirmation in science. In other words, an observation in the analogue gravity can confirm the observation in the original gravitational phenomenon. However, this particular stance receives critiques and supports. Supporters believe that if the physics of the analogue and the original gravitational phenomenon are the same, then, observation in one inevitably leads to the other because of the fact that the same mathematical structure governs the two phenomena. Nevertheless, the critics believe that there are some phenomena, say the Hawking radiation and the string theory, are not confirmed empirically, then on what basis we consider their mathematical structure so perfectly valid, put it differently, although the equations of motion of the analogue and the original phenomenon are the same, how do we know that the original phenomenon is described by that mathematical structure? how do we know that the real string is described by string theory? They have not been confirmed empirically yet.

In this paper we provide an argument based on the structuralism which states that the analogue gravity can play the role of empirical confirmation if it can simulate the original phenomenon. The rest of the paper is organized as follows: a brief introduction of physical background of analogue gravity is provided, then arguments in favor of and against the confirmatory role of the analogue gravity are considered and finally an argument based on the structuralism is pushed forward. 

%----------------------------------------------------------------------------------------
\section{Analogue Gravity}
There are some well-known correspondences in physics that relate two different theories, say an AdS gravity to a conformal field theory known as AdS/CFT [Aharony et.al, 2000], or physical systems, a gravitational phenomenon to a condensed matter system. These correspondences let us develop our understanding of the core nature of the physics, let alone they suitably equip us with some mathematical techniques to solve problems.

Analogue gravity shows that if there is a perturbation in a condensed matter system, then the equations of motion are exactly the same as the equation of motion for a field theory in a curved background [Barcelo et.al, 2011]. Suppose a condensed matter system, say a fluid\footnote{Electrons in many condensed matter systems behave as a fluid.}, Euler and continuity equations for a vortex free fluid are as follows [Barcelo et.al, 2011]:
\begin{align}
&\nabla \times \vec{v}=0\\
&\rho \big(\frac{\partial \vec{v}}{\partial t}+(\vec{v}. \nabla)\vec{v}\big)=-\nabla p-\rho \nabla V \label{euler}\\
&\frac{\partial \rho}{\partial t}+\nabla . (\rho \vec{v})=0\\
\end{align}
where $\vec{v}$ is the velocity of the fluid, $\rho$, the density, $V$, potential and $p$ the pressure. Changing variables:
\begin{align}
&\xi:=ln \rho~~~\vec{v}:=\nabla \psi\\
&g(\xi):=\int^{e^\xi}\frac{1}{\rho'}\frac{dp(\rho')}{d\rho'}d\rho'
\end{align}
Finding linearized equations:
\begin{align}
&\xi=\xi_0+\bar{\xi}~~~\psi=\psi_0+\bar{\psi}\\
&\frac{1}{\rho_0}\Big(\frac{\partial}{\partial t}\frac{\rho_0}{g'(\xi_0)}\frac{\partial \bar{\psi}}{\partial t}+\frac{\partial}{\partial t}\frac{\rho_0 \vec{v_0}}{g'(\xi_0)}. \nabla \bar{\psi}+\nabla .\big(\frac{\rho_0 \vec{v}}{g'(\xi_0)}\frac{\partial \bar{\psi}}{\partial t}\big)-\nabla .\rho_0 \nabla \bar{\psi}+\nabla .\big(\frac{\vec{v} \rho_0 \vec{v}. \nabla \bar{\psi}}{g'(\xi_0)}\big)\Big)=0\label{fluideq}\\
\end{align}
then \eqref{fluideq} could be rewritten as follows:
\begin{align}
&\partial_{\mu}\big(\sqrt{-g} g^{\mu \nu}\partial_{\nu}\bar{\psi}\big)=0\label{qfcb}
\end{align}
where the metric is given by:
\begin{align}
&ds^2=\frac{\rho_0}{c(\rho_0)}\Big(\big(c^2(\rho_0)-v^2\big)dt^2+2dt \vec{v_0}. d\vec{v}-dx^2\Big) \label{line}\\
&c^2(\rho_0)=g'(ln \rho_0)
\end{align}
where $c$ is the speed of sound. In this manner, a fluid in a flat background\footnote{There is a generalization of the flat to curved background [Hu, 2019].} \eqref{euler} can mimic a field theory\footnote{For simplicity, we considered a scalar field; however, it could be generalized to an arbitrary field theory.} in a curved background \eqref{qfcb} but the speed of light is replaced with the speed of sound and black hole solutions turn into dumb hole solutions where a sound wave cannot escape. The classical field theory $\psi$ or $v$ is a classical field on a curved background constructed by $\rho, p$ and $v$ where the role of the metric $g^{analogue}_{\mu\nu}$ (the line element \eqref{line}) is played by $(\rho, p, v)$. The model of the classical field theory on the curved background (CFCB) could be demonstrated as:
\begin{align}
M_S^{CFCB}=<(\rho, p,v), v>=<g^{analogue}_{\mu\nu},v>
\end{align}

Quantization of the sound waves is as the same as the quantization of the field theory in a curved background or a black hole and as a consequence, there appears the Hawking radiation-like phenomenon in the analogue gravity [Unruh, 1981]. In a fluid, this Hawking radiation is the excitations of the phonon in the fluid that has been detected in laboratories [Weinfurtner, 2011] [Steinhauer, 2016]. The model of the quantum aspect of the field theory on the curved background, in short quantum field theory on the curved background (QFCB) could be demonstrated as:
\begin{align}
M_S^{QFCB}=<(\rho, p,v)_{classical}, v_{quantized}>=<g^{analogue}_{\mu\nu},v_{quantized}>
\end{align}
where $v_{quantized}$ is the quantization of the classical field theory $v$ or $\psi$. The main question is that if the detection of this Hawking radiation-like phenomenon in a fluid can be accounted as an empirical confirmation of the original Hawking radiation or not.
%-----------------------------------------------------------------------------------------
\section{Arguments in Favor/Against of Confirmatory Role of AG}
In science, many scientists invoke analogies to confirm their claims or at least guide them to discoveries. In this vein, the analogue gravity by trading on syntactic isomorphism between the analogue system and the original which (is inaccessible) is going to be empirically confirmatory [Dardashti, et al 2017]. A rudimentary example is the correspondence between the oscillation of a pendulum and the oscillation of electric charges in a system of RLC circuit. In both cases, the equations of motion are the same, like what happens with \eqref{euler}, \eqref{qfcb} and as a consequence, a syntactic isomorphism; nonetheless, the both systems have quite different ontology, on the one hand, there is a pendulum, gravity and the resistance of air, on the other, resistor, inductor and capacitor. This isomorphism lets us to measure the location of the pendulum and based on that find the the amount of electric charges passing through the cross section of the wire. The very similar case is brought when one considers a negative test charge near a uniform spherical configuration of positive electric charges and a point mass particle near a uniform spherical configuration of mass. The equations of motion for the test charge and the point mass are the same.

In [Dardashti ,et al, 2017] there is a line of reasoning that how this inference is justified in the analogue gravity. Suppose there is a target system $T$ which is modeled by $M_T$ under conditions $D_T$ \footnote{As van Frasen notes $T$ is a data model and $M_T$ is the theory model, that means $M_T$ represents $T$ the data model which is the result of experimental observations, and $T$ is the representation of the natural phenomenon, roughly speaking when talking of $T$ it means the natural phenomenon [van Frasen, 2006].} and system $S$ simulates the target which is modeled by $M_S$ under the conditions $D_S$:

\begin{enumerate}
\item There is a mathematical similarities between $M_S$ and $M_T$ sufficient for a syntactic isomorphism under the conditions $D_S$ and $D_T$.
\item System $T$ is inaccessible within the conditions $D_T$.
\item System $S$ within the conditions $D_S$ is accessible and we can form a claim such as: a phenomenon $P_S$ is exhibited under the conditions $D_S$ in the system $S$.
\end{enumerate}
the isomorphism lets us to infer the following statement from the above assumptions:
\begin{itemize}
\item Claim: under conditions $D_T$ the system $T$ exhibits a phenomenon $P_T$.
\end{itemize}
To be assured that the model $M_T$ describes $T$ accurately enough, it presupposes that there are some shared implicit assumptions between $M_S$ and $M_T$ where the phenomena $P_S$ and $P_T$ are based on these implicit assumptions. In addition, these implicit assumptions relate the two systems. What makes these implicit assumptions different are model-external basis and the empirically grounded. In other words, if a universality of a phenomenon is established there would be a model-external basis for the phenomenon and if it is empirically tested then it is not merely a theoretical curiosity, it is empirically grounded, what Dardashti et.al called MEEGA (model-external and empirically grounded argument). [Dardashti et.al, 2017] claims that Hawking radiation is appeared in different analogue systems and they have been empirically confirmed then they have concluded:

\begin{itemize}
\item There are shared implicit assumptions between the analogue gravity and the original gravitational phenomenon.
\end{itemize}

[Dardashti et.al ,2017] emphasizes that this argument is different from the analogical reasoning. In analogical reasoning by finding a correspondence ( similarity) between two objects one develops the correspondence and generalizes it to the other properties not included in the primer correspondence[Bartha,2019]. For example, a flower and a telecommunication tower are similar in their configuration, a tall construction and a big head. However, in analogical reasoning one develops this correspondence and concludes that if the flower has a property, say have a leaf then there would be the analogous property, say an analogous leaf in the tower. Sometimes it works and the tower has some wave transmitter as analogous leaf and sometimes it does not work. The point is that the counter examples of the analogical reasoning arise from the development of the correspondence and it is worth mentioning that the analogue gravity works within the domain of the correspondence and does not develop it. Then as long as one works in the domain of the correspondence and does not develop it, there is no problem with the analogical reasoning. Put it differently, as long as one takes the equations of motion of the analogue gravity and the original Hawking radiation and their quantizations, the correspondence is concrete and legitimate. However, if one broadens the correspondence, say to the singularity of solutions, the analogy does not work because of the fact that the analogue solutions are not singular but the real black hole solutions.

There remains a big question: whether the model $M_T$ and $M_S$ are empirically adequate enough or not? Obviously, the model $M_S$ which is accessible is easy to check in experiments; however, the knotty part of the problem lays in the accuracy and the adequacy of the model $M_T$ which is inaccessible. It seems that there is not a clue for the accuracy of the original Hawking radiation and actually it is a vicious circle. Put it another way [Crowther, 2021]:
\begin{itemize}
\item The model describing the original Hawking radiation is not confirmed experimentally and consequently not empirically adequate.
\item The analogue gravity is going to provide the experimental confirmation.
\item To do that one should make sure that the model of the original Hawking radiation is adequate enough.
\end{itemize}
[Dardashti et.al, 2019] tries to invoke a probabilistic analysis to ascertain that the more phenomena following the Hawking radiation, the more plausibility of the original Hawking radiation. Having said that, this analysis makes sure that if the model $M_T$ is partially true then more experimental evidences in the analogue gravity increases the accuracy of $M_T$; nevertheless, this conclusion still tapped in the previous vicious circle. In other words, the adequacy of the Hawking radiation model is still under  question [Crowther et.al, 2021].

Some philosophers of physics call for the inductive inference [Evans, 2020].  Evans et.al claim that most of modern physics is due to unobservable phenomena; however, some of them are manipulable such as spin of the electrons and some of them are unmanipulable such as cosmological phenomena. Unmanipulable ones can be divided into accessible (indirectly observed) such as astrophysical phenomena and inaccessible. They claim that in these areas of research physicists carry out their research via inductive reasoning and if one is not that skeptic about inductive reasoning. [Evans, 2020] provides an interesting example of interstellar nucleosynthesis, where scientists talk about the nuclear reactions in the core of the stars in unmanipulable and inaccessible position, so scientists find that if they make some assumptions ( the assumptions are: the atomic theory (manipulable and accessible), photonic spectra of the surface of the star (unmanipulable and accessible) ), they can successfully describe the energy emitted from stars\footnote{Although Evans et.al using the inductive inference in the nucleosynthesis argument, it seems that the proper reasoning is IBE (inference to the best explanation). We talk on this later.}. [Evans et.al, 2020] calls this inductive reasoning as inductive triangulation, combination of two inductive inferences comes to a conclusion. Then this kind of inference can be applied to the analogue gravity and conclude that the observation of the Hawking radiation in the analogue gravity leads to the empirical confirmation of the original Hawking radiation which are unmanipulable and inaccessible. 

It seems that [Evans et.al, 2020] calls for the inductive inference to show that it is reasonable to suppose a particular nuclear reaction for the deep interior of stars; however, the main question to answer is that if the two assumptions, the atomic theory plus the surface photons of the star, imply the empirical confirmation of the interior nuclear reaction or not. We think that they do answer the different question. The question they are answering is if it is reasonable to describe interior of the stars based on other theories? They do not tell anything about the fact that whether this description is empirically adequate. In addition, [Evans et.al, 2020] truly states that in the inductive inference, systems $S$ and $T$ pertain to the same system; however, in the analogue gravity $S$ and $T$ are in the same universality class \footnote{Evans et.al emphasize that if we suppose that $T$ is accurately described by $M_T$, $M_T$ and $M_S$ are isomorphic.}. 

To demonstrate that the analogy in Hawking radiation is not legitimate Field draws the relation between the universality and empirical constraints in theories based on distinguishing between two factors: strength and relevance [Field, 2021]. Strength shows the possibility of the conclusion from the premises and the relevance concerns how the universality is positively relevant to the system of interest. Field addresses the relevance to the microscopic shared structure of some superficially different macroscopic theories such as renormalization group theory in condensed matter systems with the shared microscopic mathematical structure but different macroscopic theories of condensed matter systems. For inaccessible systems we can make a guess and see how well our guess works in observation if the guess is relevant or not; however, in Hawking  radiation we cannot follow this procedure due to the inaccessibility. In addition, Field states that Hawking radiation and the analogue gravity due to the difference of microscopic structure are not relevant, then we are not allowed to use of Hawking radiation in analogue models. 

We raise an objection to the Field's exposition. In the correspondence between RLC circuit and the pendulum, they share no microscopic mathematical structure and only share the same macroscopy. However, one is hundred percent sure that if measuring a quantity in the circuit there would be absolutely an analogous quantity in the pendulum system if measuring it one gets the same magnitude of its analogous circuit. In this way, it seems that the correspondence between the original Hawking radiation and its analogue is reasonable due to their shared macroscopic structure, with different microscopic structures though. In other words, if there is Hawking radiation in the analogue gravity then the shared macroscopic mathematical structure makes sure that the original Hawking radiation occurs unless there is a doubt that if this shared macroscopic mathematical structure does describe the original black hole correctly.
%------------------------------------------
\subsection{Structuralism}
It appears that if we make sure that $T$ is adequately described by $M_T$ and the models $M_T$ and $M_S$ are isomorphic then the observation $P_S$ can provide an empirical confirmation for $P_T$; however, the main problem in here is around the question: on what basis one can make sure that an unmanipulable and inaccessible system such as the original Hawking radiation, $T$, is described by the Hawking's calculations, $M_T$ without empirical confirmation.

Inference to the best explanation (IBE) is spelled out where there are competing assumptions and the all are empirically adequate, the assumption that gives the best explanation is true \footnote{There are some critiques to this kind of inference that we do not consider them here and take IBE for granted. For a latest review see [Douven, 2021].} [Ladyman, 2001]. Hawking's calculations are based on two assumptions: quantum field theory and general theory of relativity. Both theories are empirically adequate; however, in Hawking's calculations a field theory in a curved background is considered, it means a combination of the both theories. These calculations lead to a temperature for black holes. Based on IBE, it is reasonable to generalize the quantum field theory to the curved background as it happened in generalization of special theory of relativity to  general theory of relativity. In this manner, IBE provides a justification of the Hawking's calculations due to the current evidences we have got in hand from general theory of relativity and quantum field theory similar to argument in [Evans et.al, 2020] about nucleosynthesis. However, it seems that if the next radical scientific change predicts the violation of the Hawking's calculation, then all of the analogue gravity program of empirical confirmation is jeopardized.

Nevertheless, structuralism seems to be promising particularly when considering the explicit reconstruction of theories from empirical science, structuralism is distinguished between the rivals [Balzer et.al, 2000] [Niebergall, 2002]. This school of thought committees us to the mathematical structure of the scientific theories and avoids the theory description of the furniture of the world [Ladyman, 2014]. Particularly the structuralism evades the known pessimistic meta-induction\footnote{Pessimistic meta-induction asserts that if there is a radical change in scientific theories then the best current established scientific theories are abandoned [Ladyman, 2001].}, uderdetermination of scientific theories \footnote{It is not guaranteed that the evidences determine a unique theory which means that there is an underdetermination problem in scientific theories.} [Lee, 2022] and it supports the Putnam's no-miracle argument \footnote{No-miracle argument is mostly one of the compelling argument in favor of the scientific realism which states that there would be a miracle if the success of the scientific theories were not at least approximately true description of the world [Ladyman, 2001].} \footnote{It is worth mentioning that there are some criticism on the structural realism such as the problem of higher structures [Roberts, 2010]; however, these are due to the metaphysical status of the structures in the structural realism which are evaded in the structuralism without metaphysical status of the structures. By structuralism in this paper we mean the latter one.}.

The structural continuity of the scientific theories in structuralism gives rise to the fact that the structure of the new theory reduces to the structure of the old theory under the old theory condition. For example, it has been shown that the structure of quantum mechanics reduces to the classical mechanics [Saunders, 1993], special relativity to classical mechanics [Brown, 1993], general theory of relativity to the Newtonian gravitation [Fletcher, 2019] and general theory of relativity to geometrized Newtonian gravitation [Masoumi, 2021]. 

What we mean by structure is the meaning of the structure in the semantic approach. It is believed that the semantic approach to scientific theories is more appropriate when philosophers of science considering the practice of scientists. In this sense, a theory is a family of models rather than axiomatic systems. In addition, the structure is models of the theory in form of [French et.al, 2006]:
\begin{align}
M=<S,R_i, \phi_j, s_k>
\end{align}
where $S$ is a non-empty set, $R_i$ family of relations, $\phi_j$ family of functions and $s_k$ family of distinguished individuals of the set $S$. One can demonstrate an evaporating black hole by the following model:
\begin{align}
M_{T=Hawking~radiation}=<\mathcal{M}, g_{\mu\nu}, \phi>
\end{align}
where $\mathcal{M}$ is a manifold, $g_{\mu\nu}$ a metric (the agent of general theory of relativity) and $\phi$ a field theory (the agent of field theory). If the field $\phi$ is a classical field theory, such as Maxwell field, there are a lot of experimental evidences on the classical field theory on the curved background; for example, Shapiro time delay effect, blue shift and red shift in astronomical observations, then the structure of the whatever the next scientific theory should reduce to the structure of CFCB:
\begin{align}
M_T^{new}\rightarrow M_T^{CFCB}=<\mathcal{M}, g_{\mu\nu}, \phi_{classical}>
\end{align}
Then it seems that the simulation and detection of the effects caused by CFCB in the analogue gravity is reasonable. Such experiments for the Hawking radiation have been conducted and the ratio of the modes of the classical field has been measured in a water tank experiment [Weinfurtner et.al ,2011]. But here the empirical confirmation of CFCB is independent of the analogue gravity and it is based on the direct observation. Because of the fact that CFCB is empirically adequate ( as said before, under the CFCB condition), then by structuralism these structures will appear in the whatever the next scientific theories. 

We can summarize the points as follows:
\begin{itemize}
\item There is an independent empirical confirmation of classical fields on the curved background (CFCB)\\
\item There exists an isomorphism between the structure of the analogue gravity, $M_S^{CFCB}$ and the structure of the classical Hawking radiation, $M_T^{CFCB}$ \\
\item The whatever the next scientific theory will reduce to the structure of the classical Hawking radiation:
\end{itemize}
\begin{align}
M_T^{new}\rightarrow M_T^{CFCB}\cong M_S^{CFCB}=<g_{\mu\nu}^{analogue}, v>
\end{align}

So far we have emphasized that the structure of the classical aspect of the original Hawking radiation is empirically adequate and isomorphic to that of the analogue gravity but there is no experimental evidence on the presence of a quantum field theory on a curved background (QFCB).

To show that it is reasonable to expect that the quantum aspect of the Hawking radiation should be in the whatever the next scientific theory, we restrict our attention to the early times of the black hole evaporation or the times before the so-called Page time\footnote{After the Page time there would be a paradox in the Hawking's calculations known as the information paradox. This paradox in a nutshell is that a pure state for a closed system evolves to a mixed state which is in contradiction to the unitarity of the quantum theory. Although the Hawking's calculation leads to an inconsistency after the Page time, AdS/CFT correspondence provides a resolution to the paradox which is known as the island prescription [Almhri et al., 2020]. The information paradox is conspicuous when one is going to find the entanglement entropy of the Hawking radiation. However, AdS/CFT translates the entanglement entropy of the Hawking radiation ( a calculation in a quantum field theory or the boundary theory) into the area of a surface in AdS ( a calculation in a classical theory or the bulk theory). The island prescription in which the entanglement entropy is computed as a surface in the classical theory is in agreement with the Hawking's calculation before the Page time and different after the Page time. Owing to the fact that this prescription is a computation in the gravitational theory (the bulk theory), it does not tell us how the Hawking radiation is described in the field theory. However, one can simulate the bulk theory of the Hawking radiation by the analogue gravity and correspond the information paradox and the island role to the momentum loss over the dumb hole horizon and pump maintenance of the momentum in the dumb hole, respectively [Parvizi et al., 2023]. Nonetheless, it is controversial that if the two dual theories are exactly equivalent or not ( see, [Rickles, 2017] as a proponent and [Dardashti et.al., 2018] as an opponent), let alone the confirmatory role of the two dual theories. In short, the confirmatory role of the analogue gravity after the Page time is out of the scope of this work.}. The point is that the Hawking's calculation after the Page time leads to the information paradox ( for a review see [Manchak et al,2018][Raju,2022]) which means that as long as we are ahead of the Page time this calculation is an appropriate description based on IBE. Now we argue as follows:
\begin{enumerate}[I)]
\item It is believed that the quantum theory is one of the best current fundamental physical theories.
\item The quantum description of a phenomenon where its classical description is empirically confirmed, is a more complete description of the phenomenon.
\item Then: it implies that the quantum description of the classical Hawking radiation is a more complete description and structurally developed one. As a consequence, the structure of the whatever the next scientific theory should reduce to the structure of the quantum aspect of the Hawking radiation. \label{arg}
\end{enumerate} 
which means that:

\begin{center}
\begin{tabular}{ccccc} 
$M_T^{new}$ & $\rightarrow$  & $M_T^{QFCB}$ & $\rightarrow$  & $M_T^{CFCB}$ \\ 
 &  & $\cong$ &  & $\cong$ \\
& &$M_S^{QFCB}$&$\rightarrow$&$M_S^{CFCB}$ \\

\end{tabular}
\end{center}
In the previous section it has been shown that the missing parts of the arguments in favor of the empirical confirmation via the analogue gravity are the following:
\begin{enumerate}[1)]
\item Whether real black holes are described by the Hawking's calculation.
\item If so, why one should infer the empirical confirmation of the Hawking radiation based on its analogous one in the analogue gravity.
\end{enumerate}
The reason for the first one is conclusion \ref{arg}. The reason to support the second one is the fact that the observation of the Hawking radiation in the analogue gravity means the empirical confirmation of the structure $M_S^{QFCB}$ which is isomorphic to the structure $M_T^{QFCB}$.

All in all, if one takes the stance of structuralism, the mathematical structure of the new theory reduces to the structure of the old theory and subsequently, the structure of the Hawking's calculations (before the Page time) as an old theory is retrieved. In other words, the structure of the whatever the next scientific theory, $M_T^{new}$, will be reduced to the structure of the Hawking radiation, $M_T^{QFCB}\cong M_S^{QFCB}$ where the structure of the Hawking's calculation is isomorphic to the structure of the analogue gravity. Resultantly, empirical confirmation of the Hawking radiation in the analogue gravity leads to the empirical confirmation of the structure $M_S^{QFCB}\cong M_T^{QFCB}$. Moreover, the observation of $P_S$ prompts the empirical confirmation of the original Hawking radiation,  $P_T$.

%%%%%%%%%%%%%%%%%%%%%%%%%%%%%%%%%%%%%%%%%%%%%%%%%%
\section{Conclusion}
String theory, Hawking radiation and some quantum gravitational phenomena are yet to be confirmed empirically and they are inaccessible and will be so, perhaps. Then there would be a problem for the empirical confirmation of these phenomena. Analogue gravity by simulation of these phenomena seems that provides a fertile ground for the empirical confirmatory role of these areas of research. Supporters of this idea claim that the mathematical structure of analogue gravity models and the quantum gravitational phenomena are the same. Accordingly, if a phenomenon, say, Hawking radiation occurs in the analogue gravity side, the shared mathematical structure necessitates the occurrence of the analogous phenomenon in the quantum gravity side. However, critics believe that on what basis people think that the original quantum gravitational phenomenon is described by the shared mathematical structure if it is not confirmed empirically.

Structuralism believes that the continuity and development of science are demonstrated by the continuity and development of the mathematical structure of the scientific theories. As a consequence, the mathematical structure of a new theory should be partially reduced to the mathematical structure of the old theory. In this manner, the mathematical structure of new theories about quantum gravity should be reduced to that of the old ones, say, partially to the Hawking's calculations on the Hawking radiation. Because of the fact that Hawking's calculations share the same mathematical structure with the analogue gravity then, partially, the occurrence of the Hawking radiation in the analogue gravity confirms the occurrence of the Hawking radiation in original black holes.

%------------------------------------------------
\section*{Acknowledgment}   
The author would like to acknowledge his debt to Saeed Masoumi for helpful discussion and appreciate Radin Dardashti and Paul Bartha for reading the manuscript and very useful comments. 
%----------------------------------------------------------------------

%---------------------------------------------------------+++-----------


\begin{thebibliography}{99}
\bibitem{}Aharony, O., Gubser, S., Maldacena, J., Ooguri, H., Oz, Y., 2000. "Large N Field Theories, String Theory and Gravity", Phys.Rept. 323 (2000) 183-386

\bibitem{}Almhri, A., Mahajan, R., Maldacena, J., Zhao, Y., 2020. "The Page curve of Hawking radiation from semiclassical geometry", JHEP 03 (2020), arXiv:1908.10996v2 [hep-th]

\bibitem{}Balzer, W., Moulines, C. U., Sneed, J., 2000. "Structuralist Knowledge Representation–Paradigmatic Examples", Rodopi, Amsterdam

\bibitem{}Barcelo et.al., 2011. "Analogue gravity", Living Rev.Rel. 14 (2011) 3

\bibitem{}Bartha, P., 2019. "Analogy and Analogical Reasoning", in E. N. Zalta (ed.), The Stanford Encyclopedia of Philosophy, http://plato.stanford.edu/archives/
fall2013/entries/reasoning-analogy.

\bibitem{}Brown, H.R., 1993. “Correspondence, Invariance and Heuristics in the Emergence of Special Relativity,” in S. French, and H. Kamminga (eds.), Correspondence, Invariance and Heuristics (Boston Studies in the Philosophy of Science: Volume 148), Dordrecht: Springer, 227–260.

\bibitem{}Crowther, K., Linnemann, N., Wuthrich, C., 2021. "What we cannot learn from analogue experiments", Synthese 198 (2021) Suppl 16, 3701-3726

\bibitem{}Dardashti, R., Thébault, K., Winsberg, E., 2017. "Confirmation via Analogue Simulation: What Dumb Holes Can Tell us About Gravity", The British Journal for the Philosophy of Science 68(1)

\bibitem{}Dardashti, R., Dawid, R., Gryb, S., Thébault, K., 2018.
"On the Empirical Consequences of the AdS/CFT Duality",  arXiv:1810.00949v1 [physics.hist-ph]

\bibitem{}Dardashti, R., Hartmann, S., Thébault, K., Winsberg, E., 2019. "Hawking Radiation and Analogue Experiments: A Bayesian Analysis", Stud.Hist.Phil.Sci.B 67 (2019) 1-11

\bibitem{}Douven, I., 2021. "Abduction", The Stanford Encyclopedia of Philosophy (Summer 2021 Edition), Edward N. Zalta (ed.), URL = https://plato.stanford.edu/archives/sum2021/entries/abduction

\bibitem{}Evans, P. W., Thébault, K., 2020. "On the limits of experimental knowledge", Philosophical Transactions of the Royal Society A, Volume 378, Issue 2177, article id.20190235

\bibitem{}Field, G., 2021. "Putting theory in its place: The relationship between universality arguments and empirical constraints", BJPS

\bibitem{}Fletcher, S.C., 2019. "On the reduction of general relativity to Newtonian gravitation", 
Studies in History and Philosophy of Science Part B: Studies in History and Philosophy of Modern Physics, Volume 68, 2019, Pages 1-15

\bibitem{}French, S., Ladyman, J, 2006. "Reinflating the semantic approach", International studies in the philosophy of science 13 (2), 103-121

\bibitem{}Manchak, J. B.,  Weatherall, J. O., 2018. "(Information) Paradox Regained? A Brief Comment on Maudlin on Black Hole Information Loss", Foundations of Physics, 48(6), 611-627, arXiv:1801.05923v2 [physics.hist-ph]

\bibitem{}Raju, S., 2022. "Lessons from the Information Paradox", Phys.Rept. 943 (2022) 1-80, arXiv:2012.05770v2 [hep-th]


\bibitem{}Ladyman, J., 2001. "Understanding Philosophy of Science",  Routledge (December 20, 2001)

\bibitem{}Ladyman, J., 2014. "Structural Realism", The Stanford Encyclopedia of Philosophy (Winter 2020 Edition), Edward N. Zalta (ed.), URL = https://plato.stanford.edu/archives/win2020/entries/structural-realism

\bibitem{}Lee, C., 2022. "The structuralist approach to underdetermination", Synthese volume 200, Article number: 108 (2022)

\bibitem{}Masoumi, S., 2021. "On the Continuity of Geometrized Newtonian Gravitation and General Relativity", 
Foundations of Physics 51 (2): 1-33. 2021

\bibitem{}Niebergall, K.G., 2002. "Structuralism, Model Theory and Reduction", Synthese Vol. 130, No. 1, Structuralism (Jan., 2002), pp. 135-162 (28 pages)

\bibitem{}Parvizi, S., Shahbazi, M., 2023. "Analogue gravity and the island prescription", arXiv:2302.08742 [hep-th]

\bibitem{}Rickles, D., 2017. "Dual theories: ‘Same but different’ or ‘different but same’?", Studies in History and Philosophy of Science Part B: Studies in History and Philosophy of Modern Physics 59:62-67 (2017)

\bibitem{}Roberts, B.W., 2010. "Group Structural Realism", The British Journal for the Philosophy of Science, 61 (4).

\bibitem{}Saunders, S., 1993. “To what physics corresponds,” in S. French and H. Kamminga (eds.), Correspondence, Invariance and Heuristics: Essays in Honour Of Heinz Post (Boston Studies in the Philosophy of Science: Volume 148), pp. 295–325. Dordrecht: Kluwer Academic Press.


\bibitem{}Steinhauer, J., 2016. "Observation of quantum Hawking radiation and its entanglement in an analogue black hole", Nature Physics
volume 12, pages959965 (2016)


\bibitem{}Unruh, W.G., 1981. "Experimental Black-Hole Evaporation?", Phys. Rev. Lett. 46, 1351

\bibitem{}van Fraassen, B. C., 2006. "Representation: The Problem for Structuralism", Philosophy of Science Vol. 73, No. 5

\bibitem{}Weinfurtner, S., Tedford, E.W., Penrice, M.C.J., Unruh, W.G., Lawrence, G.A., 2011. "Measurement of stimulated Hawking emission in an analogue system", Phys.Rev.Lett.106:021302,2011


	
\end{thebibliography}
\end{document}